\documentclass[preprint,showpacs,amsmath,amssymb,aps]{revtex4-1}

\usepackage{graphicx}

\def\be{\begin{equation}}
\def\ee{\end{equation}}
\def\bea{\begin{eqnarray}}
\def\eea{\end{eqnarray}}

\def\Cp{{C_p}}
\def\Va{{V_a}}
\def\kT{{k_BT}}

\begin{document}

\title{Unified derivation of phase-field models for alloy 
solidification from a grand-potential functional}

\author{Mathis Plapp}
\affiliation{Physique de la Mati\`ere Condens\'ee, 
\'Ecole Polytechnique, CNRS, 91128 Palaiseau, France}

\date{\today}

\begin{abstract}
In the literature, two quite different phase-field formulations
for the problem of alloy solidification can be found. In the
first, the material in the diffuse interfaces is assumed to be
in an intermediate state between solid and liquid, with a unique
local composition. In the second, the interface is seen as a 
mixture of two phases that each retain their macroscopic 
properties, and a separate concentration field for each 
phase is introduced. It is shown here that both types of models 
can be obtained by the standard variational procedure if
a grand-potential functional is used as a starting point
instead of a free-energy functional. The dynamical variable
is then the chemical potential instead of the composition.
In this framework, a complete analogy with phase-field models
for the solidification of a pure substance can be established.
This analogy is then exploited to formulate quantitative phase-field
models for alloys with arbitrary phase diagrams. The precision
of the method is illustrated by numerical simulations with
varying interface thickness.
\end{abstract}

\pacs{64.70.Dv,81.30.Fb,05.70.Ln,81.10.Aj}

\maketitle

\section{Introduction}

The development of the phase-field method has led to
tremendous progress in the modelling of pattern formation
during solidification, due to its capability to
simulate complex time-dependent and three-dimensional
morphologies with relatively simple numerical 
codes \cite{Karma98,Boettinger02,Plapp07,Steinbach09}.
The general principle of this method is to describe 
a multi-phase system by a set of phase fields which take
constant values in each of the bulk phases and vary
smoothly through interfaces of a characteristic thickness $W$.
The equations of motion for the phase fields and their
coupling to the local thermodynamic state variables 
(temperature, density, composition etc.) can be obtained,
following the basic principles of out-of-equilibrium thermodynamics, 
by taking a variational derivative of a free energy functional,
which is generally of the Ginzburg-Landau type. Mean-field 
approximations can be used to relate the parameters that appear 
in this functional to microscopic quantities, and the phase
fields can often be interpreted as order parameters.

Generally, the equations that result from the straightforward 
application of these principles are not suitable for obtaining 
quantitatively accurate simulation results on solidification 
microstructures. The reason is that the characteristic natural
thickness of the diffuse solid-liquid interfaces is a few times 
the interatomic distance, whereas solidification patterns 
typically exhibit length scales ranging from $1$ to $100$ $\mu$m.
Even with the help of modern computers and multi-scale
algorithms, both of these scales cannot be resolved at
the same time. Therefore, in order to simulate solidification
microstructures, the thickness of the diffuse interfaces in the 
phase-field model has to be artificially enlarged, sometimes 
by two or three orders of magnitude. Quantitative results
can only be expected if both the equilibrium and kinetic properties
of the interfaces remain unchanged under this procedure.

To achieve this goal, it is helpful to adopt a phenomenological 
point of view: the phase field is seen as a smoothed indicator
function (as opposed to a physical order parameter or density), 
and all equilibrium quantities and transport coefficients 
are interpolated between the phases with smooth functions of 
the phase fields that can be freely chosen. This freedom can be 
exploited to construct phase-field models with special properties. 
In particular, rescaling of the interface thickness is greatly 
simplified in models where bulk thermodynamics and interfacial 
properties can be controlled separately.

The phase-field models for the solidification of a pure
substance published in the literature are all quite 
similar \cite{LangerPF,Caginalp89,Kobayashi93,Karma98},
in the sense that they use the same set of fundamental
fields (a phase field and the temperature field), and that
the structure of the equations is the same. One reason for
this universality is that a simple and intuitive formulation
of the model in terms of these fields yields indeed, as
will be detailed below, a model in which this separation
is achieved. Therefore, the development of the ``thin interface 
limit'' \cite{Karma96,Karma98} which has paved the way to quantitative 
simulations of dendrites \cite{Karma98,Karma00} did not require a 
change in the model formulation. 

The situation is more complicated for alloy phase-field models. 
Two different approaches with quite distinct philosophies 
have been pursued in parallel. The first, which will be 
called the ``coarse-graining'' 
approach in the following, generalizes the pure substance
model by introducing a concentration field in addition to
the temperature field and by writing down a free energy
functional that depends on the phase field, the temperature,
and the concentration \cite{Wheeler92,Caginalp93,Warren95}. 
The local values of these three 
fields are -- in principle -- the coarse-grained
counterparts of the microscopic structural order parameter,
temperature, and concentration fields, and the interface
is seen as a narrow region in space where all of these
quantities can exhibit rapid spatial variations. In
contrast, the second approach, called ``two-phase''
approach in the following, treats the interfaces as a mixture 
of two phases, each of which retains its bulk properties
even inside the interface \cite{Tiaden98,Kim99}. In this approach, the 
phase field represents the volume fraction of one of the phases.
Moreover, a separate concentration field is defined for each 
phase, and the physical concentration field is obtained by
a weighted average depending on the value of the phase 
field. The introduction of two separate concentration
fields adds a supplementary degree of freedom which has to 
be removed from the problem; this is done either by a specific
partition relation \cite{Tiaden98} or by the condition
of local equilibrium between the coexisting phases \cite{Kim99}.

In the models of the ``coarse-graining'' type, the model
structure generally leads to an intrinsic coupling between bulk 
and interface properties, which makes simulation results dependent
on the chosen interface thickness. Only for the case of dilute 
binary alloys, a specific interpolation of the thermodynamic properties
through the interface has been developed \cite{Echebarria04,Ramirez04} 
which overcomes this constraint and makes quantitative simulations 
possible. In the two-phase approach, which is more phenomonological
from the outset, bulk and interfacial properties are
decoupled by construction. However, the removal of the
extra degree of freedom introduced by the model formulation
generally requires the solution of a nonlinear equation in each point 
of the interface, and thus adds significant computational complexity.

The purpose of the present paper is to show that the 
coarse-graining approach can be easily extended to
more complex alloy systems if instead of a free energy
functional a grand-potential functional is used to 
generate the equations of motion. Furthermore, an
analysis of the resulting model shows that it is
in fact perfectly equivalent to the two-phase model, which 
offers the possibility to reinterpret and simplify the latter.
The fundamentals underlying these findings can be
stated quite simply. The motion of interfaces is
controlled by the transport of a conserved extensive quantity:
energy for a pure substance, and chemical species for isothermal
alloy solidification. In sharp-interface models, this
fact is expressed by two separate laws: a transport 
equation in the volume and a conservation law of Stefan 
type at the moving boundary. In contrast, the two-phase
equilibrium at interfaces is controlled by the intensive
quantity that is conjugate to the conserved one:
temperature for pure substances and chemical
potential for alloys. It turns out that the pure
substance model has been formulated from the start
in terms of a phase field and the intensive variable
(temperature), whereas alloy phase field models are
traditionally formulated in terms of a phase field and the 
composition, which is a density of the extensive variable (number of solute atoms).
To obtain a model for alloys that has properties analogous 
to those of the pure substance model, it is sufficient
to choose a formulation in terms of a phase field and
the chemical potential, and to switch to the appropriate
thermodynamic potential, which is the grand potential.
Of course, the fact that the quantity analogous to the
temperature is the chemical potential is well known \cite{Langer80}
and has been extensively used in sharp-interface models as well 
as in a few specific phase-field models \cite{Losert98,Plapp04,Folch05},
but its general consequences have so far not been fully
appreciated in the framework of phase-field models.

It should be mentioned that a second obstacle for
obtaining quantitative results on alloys is the strong 
contrast of the solute diffusivities between solid and 
liquid, which generates spurious solute trapping when the
interface thickness is scaled up. This problem was solved by 
the so-called antitrapping current, which was introduced 
first in the coarse-graining approach \cite{Karma01,Echebarria04},
but has also been incorporated in the two-phase model \cite{Kim07,Steinbach09}. 
Since the results of the present paper do not introduce 
major changes on this point, the results of 
Ref. \cite{Echebarria04} will be taken over without 
a detailed discussion. After a change of variables, the
model is almost identical to the model of Ref.~\cite{Echebarria04},
such that the asymptotic analysis developed there remains valid.
This will be illustrated by numerical simulations that explicitly 
test the independence of simulation results of the interface
thickness for the case of a lens-shaped phase diagram.

The remainder of the paper is structured as follows. In
Sec. \ref{sec_pure}, the standard phase-field model for 
the solidification of pure substances is reviewd for
reference. Next, the grand potential formulation is
introduced and motivated in Sec. \ref{sec_grand}, 
and illustrated by several examples in Sec. \ref{sec_examples}:
a model with two parabolic free energy functions,
a dilute alloy, and an alloy with a lens-shaped
phase diagram. The relation of this model to other
phase-field models is clarified in Sec. \ref{sec_relations},
and an example for numerical simulations is presented
in Sec. \ref{sec_numerics}.
Finally, the implications of the present findings for 
the further developments of alloy phase field
models are discussed in Sec. \ref{sec_final}.

\section{Solidification of a pure substance}
\label{sec_pure}

The minimal model of solidification is considered,
which implies some standard simplifications: 
the densities of the solid and the liquid are taken 
to be equal, and heat transport is assumed to take 
place by diffusion only. As a consequence, no 
motion of matter needs to be considered, and the 
only transported extensive quantity is heat.

Under these assumptions, the state of an inhomogeneous 
two-phase system can be completely specified on a
coarse-grained (mesoscopic) scale by two fields: 
a phase field $\phi$ which indicates the local
state (liquid or solid) of matter, and the internal 
energy density $e$. The conservation law corresponding 
to heat transport is the conservation of energy, which 
writes
\be
\partial_t e = - \vec\nabla\cdot\vec j_e.
\label{eq:heatconservation}
\ee
The heat current $\vec j_e$ is given by Fourier's law,
\be
\vec j_e = - \kappa(\phi,e)\vec\nabla T
\label{eq:heatcurrent}
\ee
where $\kappa(\phi)$ is the heat conductivity, which in
general may depend on both variables, and $T$ is the 
local temperature.

If $e$ is chosen as dynamic variable, the corresponding
thermodynamic potential is the entropy. Consequently,
the system is naturally described by an entropy functional
\be
{\cal S}[\phi,e] = \int_V s(\phi,\vec\nabla\phi,e,\vec\nabla e)
\ee
where the local entropy density $s$ depends on the fields
as well as their gradients. Using the thermodynamic 
definition of the temperature,
\be
\frac{\delta {\cal S}}{\delta e}=\frac{1}{T},
\ee
the evolution equations for the fields $e$ and $\phi$
can be written in a variational form,
\be
\partial_t \phi = M_\phi \frac{\delta {\cal S}}{\delta \phi}
\ee
\be
\partial_t e = \vec\nabla\cdot \left(\kappa(\phi)\vec\nabla T\right)
             = - \vec\nabla\cdot \left(\kappa(\phi) T^2\vec\nabla 
                    \frac{\delta {\cal S}}{\delta e}\right).
\label{eq:eeq}
\ee
The first equation expresses the local maximization of the 
entropy, which occurs at a rate given by the constant $M_\phi$;
the second is identical to the conservation law, 
Eq.~(\ref{eq:heatconservation}).

While this is a perfectly viable starting point which 
has been explored by several authors \cite{Penrose90,Wang93}, 
this formulation is rarely used in practice. Simulations are almost
always carried out with models formulated in the variables $\phi$ 
and $T$ that can be obtained from free energy functionals. 
Besides historical reasons (the first phase-field models for 
solidification were formulated in this 
language \cite{LangerPF,Fix83,Collins85}), there are also formal 
considerations which make this approach preferrable. 
The main reason to choose the intensive variable $T$ 
instead of the extensive variable $e$ is that it directly 
controls the two-phase equilibrium, which makes it easier 
and more intuitive to identify the driving forces in the
model.

This point will be illustrated by obtaining the equations
of motion from a free energy functional that is constructed
using a purely phenomenological point of view. It will be 
shown below that the standard phase-field model of 
solidification is easily obtained as a special case. 
Let $f_s(T)$ and $f_l(T)$ be the free energy densities
of pure solid and liquid, respectively, and
let the corresponding equilibrium values of the
phase field be $\phi=\pm 1$. The free energy is given by
\be
{\cal F}[\phi,T]=\int_V f(\phi,\vec\nabla\phi,T) = 
  \int_V f_{\rm int}(\phi,\vec\nabla\phi) +
  g_s(\phi) f_s(T) + (1-g_s(\phi)) f_l(T),
\label{eq:ffunctional}
\ee
where the weighting function $g_s(\phi)$ is given by
\be
g_s(\phi) = \frac{1+g(\phi)}{2},
\ee
with $g(\phi)$ a function that satisfies $g(\pm 1)=1$
and $g'(\pm 1)=0$; hence, $g_s=1$ in the solid and $g_s=0$
in the liquid. The term $f_{\rm int}$ is given by
\be
f_{\rm int} = \frac 12 \sigma (\vec\nabla\phi)^2 + H f_{\rm dw}(\phi),
\ee
where $\sigma$ and $H$ are constants of dimension energy
per unit length and energy per unit volume, respectively,
and $f_{\rm dw}(\phi)$ is a double-well function with minima 
at $\phi = \pm 1$.

The motivations for this formulation are easily understood
and are common to many phase-field models. The term
$f_{\rm int}$ creates domains where the phase field is close to its 
equilibrium values $\phi=\pm 1$ (the minima of the double-well
function), separated by diffuse interfaces.
Therefore, far from the interfaces, the free energy density 
reduces to the one of the corresponding bulk phase.
The term $f_{\rm int}$ contributes to the free energy only 
inside the interfaces; this excess free energy represents 
the surface tension. The function $g_s(\phi)$ interpolates 
between the two free energy densities through the 
diffuse interface. 

A variation of the free energy functional with respect
to the two fields $\phi$ and $T$ yields
\bea
\delta {\cal F} & = & \int_V \left\{-\sigma\vec\nabla^2 \phi + Hf'_{\rm dw}(\phi)
   + \frac{g'(\phi)}{2}\left[f_s(T)-f_l(T)\right]\right\}\delta\phi(\vec x) \nonumber \\
   & & \quad\mbox{}
   + \left\{g_s(\phi) \frac {\partial f_s(T)}{\partial T} + 
           (1-g_s(\phi)) \frac {\partial f_l(T)}{\partial T}\right\}\delta T(\vec x),
\label{eq:fvariation}
\eea
where the prime stands for derivation with respect to $\phi$.

The variation of ${\cal F}$ with respect to $\phi$ is
the driving force for the phase transition. Two features are
noteworthy: (i) the requirement that $g'(\pm 1)=0$ ensures 
that the driving force vanishes outside 
of the interfacial regions, and (ii) since, at the melting
temperature $T=T_m$, $f_s(T_m)=f_l(T_m)$, the ``thermodynamical''
part of the driving force is identically zero, independently
of the value of $\phi$. The latter property implies that the
equilibrium interface profile in $\phi$ can be calculated
from the term $f_{\rm int}$ alone. This can be shown by seeking
the equilibrium solution for a planar interface along the 
$x$ direction, which can be obtained from the condition that 
the variation of ${\cal F}$ with repect to $\phi$ must vanish.
At $T=T_m$, this condition yields
\be
-\sigma\partial_{xx} \phi + Hf'_{\rm dw}(\phi) = 0
\label{eq:intgood}
\ee
which implies that the solution of this equation is
independent of the free energies $f_s(T)$ and $f_l(T)$.
As a consequence, the surface free energy $\gamma$ 
(defined as the excess free energy due to the presence of the interface)
is also independent of the bulk free energies and is given by
\be
\gamma = I \sqrt{\sigma H} = I H W,
\label{eq:gammarelation}
\ee
where $I$ is a numerical constant that depends on the
shape of the double well function $f_{\rm dw}$, and the 
interface thickness $W$ is defined by
\be
W = \sqrt{\frac{\sigma}{H}}.
\label{eq:Wdef}
\ee
Therefore, the surface tension $\gamma$ and the interface 
thickness $W$ can be freely chosen by appropriately fixing 
the two constants $\sigma$ and $H$, independently of the
bulk properties. As stated in the introduction, the interface
properties can thus be controlled independently of the
bulk thermodynamics.

The equation of motion for the phase field is obtained
from the free energy functional by the standard variational
procedure,
\be
\partial_t \phi = - M_\phi \frac{\delta {\cal F}}{\delta \phi},
\label{eq:phiequation}
\ee
which expresses the fact that the system seeks to minimize
its local free energy at a rate which is controlled
by the constant $M_\phi$. 

To obtain an evolution equation for the temperature
field, the starting point is the observation that,
by definition, the variation of ${\cal F}$ with respect to
$T$ is equal to the negative of the local entropy density,
\be
s(T,\phi) = - \frac{\delta{\cal F}}{\delta T} = 
     - g_s(\phi) \frac {\partial f_s(T)}{\partial T} - 
           (1-g_s(\phi)) \frac {\partial f_l(T)}{\partial T}.
\label{eq:sdensity}
\ee
Since $f_{\rm int}$ was chosen independent of temperature,
$s$ is a {\em local} function of $\phi$ and $T$ (no gradients
are involved) and can be simply seen as the interpolation of
the bulk entropy densities,
\be
s(\phi,T) = g_s(\phi) s_s(T) + (1-g_s(\phi)) s_l(T).
\label{eq:sofTandphi}
\ee
The use of the thermodynamic identity $de=Tds$ (valid at constant
density) yields
\be
\partial_t e = T \partial_t s = 
   T\left(\frac{\partial s(\phi,T)}{\partial \phi}\partial_t \phi
   + \frac{\partial s(\phi,T)}{\partial T}\partial_t T\right).
\label{eq:eevolution}
\ee
Note that in writing down the second equality, it is assumed
that there is no entropy production due to local dissipation,
which is equivalent to the hypothesis that the transformations
are reversible on the mesoscopic scale of a coarse-graining cell.
Furthermore, the definition of the specific heat per unit 
volume is
\be
\Cp(\phi,T) = T \frac{\partial s(\phi,T)}{\partial T} =
 -T g_s(\phi) \frac {\partial^2 f_s(T)}{\partial T^2} - T
           (1-g_s(\phi)) \frac {\partial^2 f_l(T)}{\partial T^2}.
\label{eq:Cdef}
\ee
The equations (\ref{eq:eevolution}) and (\ref{eq:Cdef}) can be 
combined with the energy conservation law, 
Eq.~(\ref{eq:heatconservation}), where the heat conductivity
$\kappa$ now depends on the variables $\phi$ and $T$, to yield
\be
\Cp(\phi,T)\partial_t T = - T \frac{\partial s}{\partial\phi} \partial_t \phi
      + \vec\nabla\left[\kappa(\phi,T)\vec\nabla T\right],
\ee
which is the desired evolution equation for the temperature
field. This equation can be further simplified by writing
the heat conductivity as the product of the specific heat 
and the thermal diffusion coefficient $D_T$, and by using
Eq.~(\ref{eq:sofTandphi}) for the entropy density. The
result is
\be
\partial_t T = \frac{1}{\Cp(\phi,T)} \left\{
         \vec\nabla\left[\Cp(\phi,T)D_T(\phi,T)\vec\nabla T\right]
               + T[s_l(T)-s_s(T)]\frac{g'(\phi)}{2}\partial_t\phi\right\}.
\label{eq:Tequation}
\ee
Note that all quantities that appear in this equation except
for the thermal diffusivity can be obtained from the bulk free
energy densities $f_s(T)$ and $f_l(T)$.

The free energy functional used in the standard formulation
of the phase-field model can be obtained from 
Eq.~(\ref{eq:ffunctional}) by linearizing the free
energy density $f(\phi,\vec\nabla \phi,T)$ around the 
melting temperature $T_m$,
\be
f(\phi,\vec\nabla\phi,T) = f(\phi,\vec\nabla\phi,T_m) + 
\left.\frac{\partial f}{\partial T}\right|_{T_m}(T-T_m).
\ee
Using the definition of the latent heat per unit volume,
$L=T_m[s_l(T_m)-s_s(T_m)]$, as well as the fact that the 
free energies of solid and liquid are equal at $T_m$, 
$f_s(T_m)=f_l(T_m)$, this expansion yields
\be
{\cal F} = \int_V \frac 12 \sigma (\vec\nabla \phi)^2 + H f_{\rm dw}(\phi) +
   \frac{L}{2T_m}g(\phi)(T-T_m),
\label{eq:Flinear}
\ee
where the constant $[s_s(T_m)+s_l(T_m)]/2$ has been disregarded
for simplicity.

The equation of motion for the phase field is then obtained from 
this linearized functional by a variational derivative.
To obtain an equation of motion for the temperature, it is
usually assumed that the specific heat is independent of
both temperature and the phase field, $\Cp(\phi,T)\equiv \Cp$. 
Then, it is easy to ``guess'' the correct equation by
realizing that the latent heat released or consumed
during the phase transformation appears as a source 
term in the diffusion equation for the temperature,
\be
\partial_t T = \vec\nabla\left(D_T(\phi,T)\vec\nabla T\right) + 
\frac{L}{\Cp} \frac{g'(\phi)}{2}\partial_t \phi.
\label{eq:Tapprox}
\ee
This equation is indeed obtained from Eq.~(\ref{eq:Tequation})
when a constant specific heat is inserted and the approximation
$T=T_m$ is made in the second term on the right hand side.
Note that, in contrast, the correct general form could not
have been easily guessed from Eq.~(\ref{eq:Tapprox}). The
underlying reason is that the linearized free energy functional
formally yields a specific heat which is zero since all second
derivatives with respect to the temperature vanish; therefore,
thermodynamic consistency between the linearized functional,
Eqs.~(\ref{eq:Flinear}), and the evolution equation for the
temperature, Eq.~(\ref{eq:Tapprox}), has been lost. The equations
are nevertheless correct, since for the case of constant 
specific heat and constant latent heat, that is
$T[s_l(T)-s_s(T)=T_m[s_l(T_m)-s_s(T_m)]=L$, the internal
energy density and the temperature are linearly related,
\be
e(\phi,T) = \frac{e_l(T_m)+e_s(T_m)}{2} + \Cp(T-T_m)-\frac{g(\phi)}{2} L.
\label{eq:eTlinear}
\ee
Then, Eq.~(\ref{eq:Tapprox}) can be directly obtained by 
combining the time derivative of Eq.~(\ref{eq:eTlinear})
with the energy conservation law, Eq.~(\ref{eq:heatconservation}).

\section{Isothermal alloy solidification}
\label{sec_grand}
A binary alloy is a mixture of two pure substances A and B.
For simplicity, it is assumed here that the atomic volume 
$\Va$ of both pure substances and of the mixture are all the same, 
and that hence the total number density of the alloy is a constant 
and equal in solid and liquid. Then, the only new field needed is
the local composition (atomic fraction) $c$ of ``solute'' (B) atoms.
Furthermore, for constant atomic volume the chemical potentials 
of A and B atoms are not independent since removal of an
A atom implies the addition of a B atom. This means that
the only new intensive variable that needs to be considered
is the chemical potential $\mu$ of the solute atoms.

The starting point of the coarse-graining approach, as pioneered 
in Refs.~ \cite{Wheeler92,Caginalp93,Warren95}, is a free energy 
functional that depends on the variables $\phi$, $T$, and $c$.
The chemical potential is then defined as the functional derivative
of the free energy functional with respect to $c$. Since $c$ 
is dimensionless, the chemical potential obtained by this procedure 
has the dimensions of energy per unit volume. This convention obscures
the thermodynamic roles of the two variables: the relevant
extensive variable from which a density should be defined is
the number of B atoms. Therefore, the variable that is 
analogous to the internal energy density in the pure substance
model is the number density of B atoms,
\be
\rho=\frac{c}{\Va},
\label{eq:rhodef}
\ee
where $\Va$ is the atomic volume (the constant volume occupied
by one A or B atom). Then, the chemical potential defined by
\be
\mu = \frac{\delta {\cal F}}{\delta\rho} = \Va \frac{\delta {\cal F}}{\delta c}.
\label{eq:mudef}
\ee
has the dimension of an energy, as is standard in basic thermodynamics. 
Since the nature of the variables is important for establishing 
the analogy between pure substance and alloy models, this
convention for the chemical potential is adopted for the
remainder of the paper. However, since it is customary
to express free energy densities in terms of the composition 
rather than the number density, both $\rho$ and $c$ will be
used in the following for ease of presentation, keeping
in mind that the two variables are simply related by 
Eq.~(\ref{eq:rhodef}).

The number density is a conserved quantity, which implies
\be
\partial_t \rho = -\vec\nabla \cdot \vec j_\rho,
\label{eq:massconservation}
\ee
where $j_\rho$ is the mass current. For isothermal solidification,
where the only thermodynamic driving force for mass diffusion
is the gradient of the chemical potential, the mass current 
is given by
\be
\vec j_\rho = - M(\phi,T,c)\vec\nabla\mu,
\label{eq:masscurrent}
\ee
where $M(\phi,T,c)$ is the atomic mobility. Combining these
equations and the definition of $\mu$ yields the equation of 
motion for $\rho$,
\be
\partial_t \rho = \vec\nabla\cdot \left(M(\phi,T,c)
     \vec\nabla\frac{\delta {\cal F}}{\delta \rho}\right).
\label{eq:ceq}
\ee

It will now be shown that this approach generally leads to
a model in which bulk and interface properties
do not decouple. To this end, it is useful to start again
from the variation of the free energy functional, now in
the variables $\phi$ and $\rho$. Since isothermal solidification
is considered, there is no variation with respect to
temperature. In order to simplify the notations, the variable
$T$ (which becomes a simple parameter for isothermal solidification)
will be dropped from the free energy densities and the mobility
from now on. The variation of ${\cal F}$ is
\bea
\delta {\cal F} & = & \int_V \left\{-\sigma\vec\nabla^2 \phi + Hf'_{\rm dw}(\phi)
   + \frac{g'(\phi)}{2}\left[f_s(c)-f_l(c)\right]\right\}\delta\phi(\vec x) 
\nonumber \\
   & & \quad\mbox{}
   + \Va\left\{g_s(\phi) \frac {\partial f_s(c)}{\partial c} + 
           (1-g_s(\phi)) \frac {\partial f_l(c)}{\partial c}\right\}\delta \rho(\vec x).
\label{eq:fvariation2}
\eea
A crucial difference which the pure substance case is 
obvious: there is no simple argument which ensures that 
the ``thermodynamic driving force'' term proportional
to $f_s(c)-f_l(c)$ vanishes.
Indeed, for two-phase equilibrium in an alloy, both
the free energy density and the concentration vary
across the interface. The value of all these quantities
in the bulk phases at two-phase coexistence 
are obtained from two conditions:
(i) the chemical potential must be the same in both phases, 
and (ii) the grand-potential density $\omega=f-\mu \rho$ must 
also be the same. Given the curves of free energy versus 
composition, the graphical interpretation of these two 
conditions is the well-known common tangent construction.

Let us examine the consequence of these conditions for
the phase-field model outlined above. Since the two
variables -- phase field and concentration -- vary
through the interface, the equilibrium interface
profile is given by two coupled nonlinear differential
equations. One is obtained from the condition of constant 
chemical potential, which remains valid in the diffuse 
interface picture, and reads
\be
\frac{\delta{\cal F}}{\delta \rho} = 
  \Va g_s(\phi) \frac {\partial f_s(c)}{\partial c} + 
           \Va (1-g_s(\phi)) \frac {\partial f_l(c)}{\partial c} = \mu_{\rm eq}(T)
\label{eq:mucondition}
\ee
where $\mu_{\rm eq}(T)$ is the equilibrium value obtained
from the common tangent construction. This equation
defines an implicit relation between the composition
$c$ and the phase field $\phi$.

The equation for the phase field, obtained as before from 
the condition that the variation of ${\cal F}$ vanishes, is
\be
-\sigma\partial_{xx} \phi + Hf'_{\rm dw}(\phi) + 
\frac{\delta f_{\rm th}}{\delta \phi} = 0,
\ee
where
$f_{\rm th}(\phi,c) = g_s(\phi) f_s(c) + (1-g_s(\phi))f_l(c)$
denotes the ``thermodynamic part'' of the free energy density.
Obviously, this equation becomes identical to Eq.~(\ref{eq:intgood}) 
only if the third term is identically zero. The physical meaning 
of this condition can be made transparent by remarking that, since
the concentration and the phase field are not independent
variables any more under the constraint of Eq.~(\ref{eq:mucondition}),
the variation of $f_{\rm th}$ with respect to $\phi$, taking into account
the constraint of Eq.~(\ref{eq:mucondition}), is
\be
\frac{\delta f_{\rm th}}{\delta \phi} = \frac{\partial f_{\rm th}}{\partial \phi} + 
\frac{\partial f_{\rm th}}{\partial c}\frac{d c}{d\phi} = 
\frac{\partial f_{\rm th}}{\partial \phi} + \frac{\mu_{\rm eq}}{\Va}\frac{d c}{d\phi}.
\ee
Therefore, if 
\be
\frac{\delta f_{\rm th}}{\delta \phi} = \frac{d}{d\phi}\left(f_{\rm th}-\mu_{\rm eq}\rho\right)\equiv 0,
\label{eq:grandcondition}
\ee
the equation for the equilibrium phase field profile reduces 
to Eq.~(\ref{eq:intgood}); in other words, the quantity 
$f_{\rm th}-\mu_{\rm eq}\rho$ must be constant through the interface.
Far from the interfaces, where $f_{\rm int}$ does not contribute,
this quantity is equal to the grand potential density. Note
that the common tangent construction implies that the two
{\em bulk} values of the grand potential must be equal.
However, for a general choice of free energy functions,
there is no reason for this condition to be valid
{\em throughout the whole interface}. As a consequence,
the interface equation and all quantities that are obtained
from its solution (surface tension, kinetic coefficients
etc.) depend on the bulk free energy densities. For realistic
values of the interface thickness, this dependence is small,
but when the interface thickness is upscaled, large errors
can occur.

This fact has been recognized by several authors, and so
far two different strategies have been followed to cure
this problem. The first is to develop specifically designed
free energy functionals that satisfy the condition of 
Eq.~(\ref{eq:grandcondition}), but are valid only for certain 
choices of bulk free energies (see below). The second strategy 
is the one of the two-phase model \cite{Tiaden98,Kim99}, in which 
two separate concentration fields, one for each phase, are used; 
the supplementary degree of freedom is then eliminated in
such a way that Eq.~(\ref{eq:grandcondition}) is satisfied.

The new idea put forward here is that a general solution
to this problem can also be obtained in the coarse-graining
spirit (using a single concentration field) when the model is derived
from a grand-potential functional instead of a free energy
functional. Indeed, the model formulated in the variables
$\phi$ and $\rho$ is equivalent to the pure substance model
formulated in terms of $\phi$ and $e$: it has the same
variational structure [compare Eqs. (\ref{eq:eeq}) and 
(\ref{eq:ceq})], and both $\rho$ and $e$ are 
densities of extensive variables. To obtain the equivalent 
of the more successful pure substance models formulated in
the variables $\phi$ and $T$, alloy models should be
formulated in the variables $\phi$ and $\mu$; the
corresponding thermodynamic potential is the grand
potential.

grand-potential functionals have been used extensively in
classical density functional theory (see \cite{Evans79} for 
a review). In the context of phase-field models, a grand
potential functional has been introduced to study solidification
with density change \cite{Conti01,Conti03}. However, in all
the cited works, the density is retained as the fundamental
field that is used to evaluate the functional. In contrast,
the grand potential in its role as a thermodynamic potential 
depends on the chemical potential. If the goal is to have a
complete formulation of the problem in terms of the dynamical
variable $\mu$, the grand-potential functional $\Omega$ should 
be a functional of the field $\mu$. In thermodynamic equilibrium,
this field is just a constant which is equal to the thermodynamic 
equilibrium chemical potential, but out of equilibrium $\mu$ can 
depend on space and time. Therefore, the field $\rho$ that appears 
in the free energy density needs to be eliminated in favor of $\mu$.
This is simple if the values of $\rho$ and $\mu$ are related by
a local and invertible function. A free energy functional of the 
form of Eq. (\ref{eq:ffunctional}), taken with free energy densities
that depend on $T$ and $c$, is a good starting point since it 
contains no nonlocal terms in $\rho$ such as $(\vec\nabla \rho)^2$. 
Moreover, for functions $f_s(c)$ and $f_l(c)$ that are convex in $c$,
the relation between $\mu$ and $c$ is monotonous and hence invertible.
Thus, it is possible to switch from $c$ to $\mu$ as the dynamic field.
After this operation, the number density is not a fundamental free 
field any more, but is obtained as a local functional derivative of 
the grand-potential functional with respect to the local chemical
potential,
\begin{equation}
\rho = -\frac{\delta \Omega[\phi,\mu]}{\delta \mu}.
\label{eq:rhodepdef}
\end{equation}
Note that the above requirements (no square gradient
terms in $c$ and convex free energy functions) implies that
the present method cannot be applied to systems that exhibit
phase separation.

In analogy with Eq.~(\ref{eq:ffunctional}), the grand-potential
functional is
\bea
\Omega[\phi,\mu] & = & \int_V \omega(\phi,\vec\nabla\phi,\mu) \nonumber \\
   & = & 
  \int_V \omega_{\rm int}(\phi,\vec\nabla\phi) +
  g_s(\phi) \omega_s(\mu) + (1-g_s(\phi)) \omega_l(\mu),
\label{eq:gfunctional}
\eea
where $\omega_{\rm int}$ is identical to $f_{\rm int}$, and the
grand potential densities of the bulk phases are obtained
by a Legendre transform of the free energies,
\be
\omega_{\nu}(\mu) = f_{\nu}(c)-\mu \rho \qquad (\nu=s,l).
\label{eq:Legendre}
\ee
This procedure can be easily performed for any convex free
energy function, either analytically or numerically. Note
that this transformation implicitly uses the equivalence between 
statistical ensembles (canonical and grand canonical) on the
mesoscopic scale. This is consistent with the general philosophy of
the coarse-graining approach, which assumes that thermodynamic
quantities can be defined on the scale of a coarse-graining cell.

The variation of the grand-potential functional (at constant 
temperature) is
\bea
\delta \Omega & = & \int_V \left\{-\sigma\vec\nabla^2 \phi + Hf'_{\rm dw}(\phi)
   + \frac{g'(\phi)}{2}\left[\omega_s(\mu)-\omega_l(\mu)\right]\right\}\delta\phi(\vec x) \nonumber \\
   & & \quad\mbox{}
   + \left\{g_s(\phi) \frac {\partial \omega_s(\mu)}{\partial \mu} + 
           (1-g_s(\phi)) \frac {\partial \omega_l(\mu)}{\partial \mu}\right\}\delta \mu(\vec x).
\label{eq:gvariation}
\eea
Since, at solid-liquid coexistence, $\omega_s=\omega_l$,
the equilibrium interface equation obtained from the
condition of vanishing variation with respect to $\phi$
is identical to Eq.~(\ref{eq:intgood}), as desired.

Furthermore, using Eq.~(\ref{eq:rhodepdef}),
the variation of $\Omega$ with respect to $\mu$ yields
\be
\rho(\phi,\mu) = - \frac{\delta\Omega}{\delta \mu} = 
     - g_s(\phi) \frac {\partial \omega_s(\mu)}{\partial \mu} - 
           (1-g_s(\phi)) \frac {\partial \omega_l(\mu)}{\partial \mu},
\label{eq:rhoofmu}
\ee
which is the equivalent of Eq.~(\ref{eq:sdensity}) for the
entropy density. Note that this can also be rewritten as
\be
\rho(\phi,T,\mu) = g_s(\phi) \rho_s(T,\mu) + (1-g_s(\phi)) \rho_l(T,\mu)
\label{eq:rhoofmuandphi}
\ee
with $\rho_\nu = \partial\omega_\nu/\partial\mu$ ($\nu=s,l$).
It is useful to restate this equation in terms of $c$ for
future use,
\be
c(\phi,\mu) = \Va\rho(\phi,\mu) = 
   g_s(\phi) c_s(\mu) + (1-g_s(\phi)) c_l(\mu),
\label{eq:cofmuandphi}
\ee
where obviously $c_{\nu}(\mu) = \Va\partial \omega_{\nu}/\partial\mu$.
Two-phase coexistence is characterized by a constant chemical potential,
$\mu=\mu_{\rm eq}(T)$; the corresponding composition profile through a
solid-liquid interface is
\be
c_{\rm eq}(\phi) = c(\phi,\mu_{\rm eq}(T)) = g_s(\phi) c_s^{\rm eq}(T) + (1-g_s(\phi))c_l^{\rm eq}(T).
\label{eq:cprofile}
\ee

The equations of motion for $\phi$ and $\mu$ are now formulated
following the same steps as for the pure substance model.
The phase field evolves toward a minimum of the grand potential,
\be
\partial_t\phi = -M_\phi\frac{\delta\Omega}{\delta\phi}
   = M_\phi\left[\sigma\vec\nabla^2\phi-Hf'_{\rm dw}
   -\frac{g'(\phi)}{2}(\omega_s-\omega_l)\right];
\label{eq:phievolution}
\ee
this equation shows that the thermodynamic driving force
for the phase transition is the difference in grand potential
densities. The evolution equation for the chemical potential
is obtained by taking the time derivative of 
Eq.~(\ref{eq:rhoofmuandphi}), which yields
\be
\partial_t \rho = 
   \left(\frac{\partial \rho(\phi,\mu)}{\partial \phi}\partial_t \phi
   + \frac{\partial \rho(\phi,\mu)}{\partial \mu}\partial_t \mu\right).
\label{eq:rhoevolution}
\ee
It is useful to define the quantity
\be
\chi(\phi,\mu)=\frac{\partial\rho(\phi,\mu)}{\partial\mu}=
  g_s(\phi) \frac{\partial\rho_s(\mu)}{\partial\mu} + 
  (1-g_s(\phi)) \frac{\rho_l(\mu)}{\partial\mu},
\label{eq:chidef}
\ee
which will play a role similar to the specific heat
in the pure substance model.
The symbol $\chi$ is chosen here because this quantity can 
be seen as a generalized susceptibility \cite{LangerBegRohu}.
Furthermore, the mobility can be written as the product of
$\chi(\phi,\mu)$ and a solute diffusion coefficient 
$D(\phi,\mu)$. Indeed, in the bulk phases, for monotonous 
(and hence invertible) functions $\rho_{s,l}(\mu)$,
\be 
\chi_\nu(\mu)= \frac{\partial\rho_{\nu}(\mu)}{\partial\mu}=
\frac{1}{\partial\mu_{\nu}(\rho)/\partial\rho}=
\frac{1}{\Va^2\partial^2 f_{\nu}(c)/\partial c^2},
\label{eq:Darken}
\ee
which is the well-known thermodynamic factor (Darken factor \cite{Darken48}).
The combination of these definitions and Eq.~(\ref{eq:rhoevolution})
with the mass conservation law, Eq.~(\ref{eq:massconservation})
yields
\be
\partial_t\mu = \frac{1}{\chi(\phi,\mu)}
   \left\{\vec\nabla\cdot\left[D(\phi,\mu)\chi(\phi,\mu)\vec\nabla\mu\right] -
      \frac{g'(\phi)}{2}\left[\rho_s(\mu)-\rho_l(\mu)\right]\partial_t
      \phi\right\},
\label{eq:muevolution}
\ee
an equation completely equivalent to Eq.~(\ref{eq:Tequation})
for the temperature in the pure substance model.

\section{Examples}
\label{sec_examples}
The model is completely specified for {\em any} set of free
energy functions for solid and liquid by the definitions
of the grand-potential functional, Eqs.~(\ref{eq:gfunctional})
and (\ref{eq:Legendre}), and the evolution equations
(\ref{eq:phievolution}) and (\ref{eq:muevolution}).
In order to illustrate some of its properties,
it is useful to work out several explicit examples. 
First, it is shown that the equivalent of the linearized
pure substance model is obtained from parabolic free energies.
Next, it will be shown that the dilute alloy model of 
Ref.~\cite{Echebarria04} can be recovered using this formalism. 
Finally, the more general case of an ideal solution model 
will be treated.

\subsection{Parabolic free energies}

The simplest phenomenological approximation for free energy
functions for fixed equilibrium compositions $c_s^{\rm eq}$ and
$c_l^{\rm eq}$ at some temperature $T$ are two parabolas,
\be
f_{\nu}(c) = \frac 12 \epsilon_{\nu} \left(c-c_{\nu}^{\rm eq}\right)^2, \qquad(\nu=s,l)
\ee
where $\epsilon_s$ and $\epsilon_l$ are constants with dimension
energy per unit volume. The chemical potential in each phase is
\be
\mu=\frac{\partial f_{\nu}}{\partial\rho} = \Va\epsilon_{\nu}(c-c_{\nu}),
\label{eq:muofc}
\ee
which can of course be inverted to yield $c$ as a function of
$\mu$ in each phase,
\be
c= \frac{\mu}{\Va\epsilon_{\nu}}+c_{\nu}^{\rm eq}.
\label{eq:cofmu}
\ee
The grand potential densities are then obtained from the Legendre
transform, $\omega_{\nu}=f_{\nu}-\mu\rho$, where Eq.~(\ref{eq:cofmu})
is used to switch variables from $c$ to $\mu$,
\be
\omega_{\nu}(\mu) = -\frac 12 \frac{\mu^2}{\Va^2\epsilon_{\nu}}-\frac{\mu}{\Va} c_{\nu}^{\rm eq}.
\ee
Of course, the use of the definition $\rho = -\partial\omega/\partial\mu$
together with $c=\Va\rho$ yields again Eq.~(\ref{eq:cofmu}). The
equilibrium chemical potential for two-phase coexistence is obtained 
by the condition $\omega_s(\mu_{\rm eq})=\omega_l(\mu_{\rm eq})$. The
solution $\mu_{\rm eq}=0$ corresponds to the common tangent between
the bottoms of the parabolic wells, and the equilibrium compositions
are equal to $c_s^{\rm eq}$ and $c_l^{\rm eq}$. For $\epsilon_s\neq\epsilon_l$,
a second solution exists which corresponds to a common tangent
that is tilted and yields different values for the equilibrium
compositions; this solution is not of interest here. 
Since $\mu_{\rm eq}=0$, we have $\mu=\mu-\mu_{\rm eq}$,
that is, in this model $\mu$ is directly the deviation from the 
equilibrium value of the chemical potential.

Inserting these grand potential densities in Eqs.~(\ref{eq:cofmuandphi}) 
and (\ref{eq:chidef}) yields the expressions for $c$ and $\chi$ as
a function of $\phi$ and $\mu$,
\be
c(\phi,\mu) = \left[\frac{1}{\Va\epsilon_s}g_s(\phi)+
                    \frac{1}{\Va\epsilon_l}(1-g_s(\phi))\right]\mu
              +c_s^{\rm eq}g_s(\phi)+c_l^{\rm eq}(1-g_s(\phi))
\label{eq:cofmuandphiparabolas}
\ee
\be
\chi(\phi,\mu) = \frac{1}{\Va^2\epsilon_s}g_s(\phi)+\frac{1}{\Va^2\epsilon_l}(1-g_s(\phi)).
\ee
It is easy to see that when $\epsilon_s=\epsilon_l\equiv\epsilon$,
the resulting model is equivalent to the standard pure substance
model. Indeed, in this case the difference between the compositions
in the two phases is independent of $\mu$ and identical to
$\Delta c = c_l^{\rm eq}-c_s^{\rm eq}$, and the susceptibility is just
a constant, $\chi=1/(\Va^2\epsilon)$. This corresponds to the
approximations of constant latent heat and constant specific
heat, respectively. It is easily verified that 
Eq.~(\ref{eq:cofmuandphiparabolas}) becomes identical to
Eq.~(\ref{eq:eTlinear}), with $\Delta c$ and $\chi$ replacing 
$L$ and $\Cp$, respectively. This analogy has been used in
several phase-field models \cite{Losert98,Plapp04,Folch05}. 
The present formalism makes it possible to generalize this 
model and to use $\epsilon_s\neq\epsilon_l$, that is, parabolas 
with different curvatures.

\subsection{Dilute alloy}

The phase-field model for a dilute alloy of Ref.~\cite{Echebarria04}
starts from the free energy densities
\be
f_{\nu}(T,c) = f_{\nu}^A(T) + \epsilon_{\nu}c + 
\frac{\kT}{\Va} \left(c\ln c - c\right), \qquad(\nu = s,l)
\label{eq:fdensitydilute}
\ee
where $f_s^A(T)$ and $f_l^A(T)$ are the free energy
densities of pure A, $\epsilon_s$ and $\epsilon_l$ are
again constants with dimension energy per unit volume,
$k_B$ is Boltzmann's constant, and the last term in
Eq.~(\ref{eq:fdensitydilute}) is the dilute limit of
the entropy of mixing.

The calculation of the grand potential densities is
straightforward and yields
\be
\omega_{\nu} = f_{\nu}^A - \frac{\kT}{\Va}
      \exp\left(\frac{\mu-\Va\epsilon_{\nu}}{\kT}\right).
\ee
The compositions as a function of the chemical potential
are given by
\be
c_{\nu} = \exp\left(\frac{\mu-\Va\epsilon_{\nu}}{\kT}\right).
\label{eq:cofmudilute}
\ee
It is obvious that these compositions satisfy, for any
value of the chemical potential, the partition relation
\be
c_s = k c_l
\ee
with the partition coefficient $k$ given by
\be
k=\exp\left(\frac{\Va(\epsilon_l-\epsilon_s)}{\kT}\right).
\label{eq:kdef}
\ee

The equilibrium chemical potential for a given temperature
is again obtained from the condition 
$\omega_s(T,\mu_{\rm eq})=\omega_l(T,\mu_{\rm eq})$ which yields
\be
f_s^A(T)-f_l^A(T)=\frac{\kT}{\Va}\left[\exp\left(\frac{\mu_{\rm eq}-\Va\epsilon_s}{\kT}\right)
                          -\exp\left(\frac{\mu_{\rm eq}-\Va\epsilon_l}{\kT}\right)\right].
\label{eq:mueqdilute}
\ee
The left hand side can be expanded for temperatures close
to the melting temperature $T_m$; the right hand side can
be rewritten in terms of $c_l^{\rm eq}=c_l(\mu_{\rm eq})$ and the
partition coefficient. The result is
\be
\frac{L}{T_m}(T-T_m) = \frac{\kT}{\Va} c_l^{\rm eq}(k-1).
\ee
If, on the right hand side of this equation, $T$ is approximated
by $T_m$, which is justified in the limit $c\ll1$, the standard 
dilute alloy phase diagram is obtained,
\be
c_l^{\rm eq} = \frac{T_m-T}{m},
\ee
where $m=k_BT_m^2(1-k)/(\Va L)$ is the liquidus slope.

When these expressions are now used in the complete
grand-potential functional, the composition as a function 
of $\phi$ and $\mu$ becomes
\be
c(\phi,\mu) = g_s(\phi) \exp\left(\frac{\mu-\Va\epsilon_s}{\kT}\right)
               + (1-g_s(\phi)) \exp\left(\frac{\mu-\Va\epsilon_l}{\kT}\right).
\label{eq:cofphiandmudilute}
\ee
A specific feature of the dilute alloy model is that this
expression can be factorized into two functions that depend
only on $\phi$ and $\mu$, respectively. Moreover, the
latter can again be rewritten in terms of the liquid
composition, which yields
\be
c(\phi,\mu)=c_l(\mu)\left[1-(1-k)g_s(\phi)\right].
\label{eq:csimplified}
\ee
As a result, the expression for the susceptibility is
also quite simple and reads
\be
\chi(\phi,\mu) = \frac{1}{\Va\kT}
  \exp\left(\frac{\mu-\Va\epsilon_l}{\kT}\right)\left[1-(1-k)g_s(\phi)\right]
    = \frac{c(\phi,\mu)}{\Va\kT}.
\label{eq:chidilute}
\ee

Inserting these expressions in Eqs.~(\ref{eq:phievolution}) and
(\ref{eq:muevolution}) leads to model equations identical
to those of Ref.~\cite{Echebarria04}. The difference of the
grand potential densities is
\begin{eqnarray}
 \omega_s(\mu)-\omega_l(\mu) & = & f_s^A(T) - f_l^A(T) \nonumber\\
  & & \mbox{} - \frac{\kT}{\Va}
   \left[\exp\left(\frac{\mu-\Va\epsilon_s}{\kT}\right)
        -\exp\left(\frac{\mu-\Va\epsilon_l}{\kT}\right)\right].
\end{eqnarray}
The free energy difference can be expressed in terms of
the equilibrium chemical potential using Eq.~(\ref{eq:mueqdilute}),
and after some algebra one obtains
\be
\omega_s(\mu)-\omega_l(\mu) = \frac{\kT}{\Va} c_l^{\rm eq}(1-k)
   \left[\exp\left(\frac{\mu-\mu_{\rm eq}}{\kT}\right)-1\right].
\ee
As in Ref.~\cite{Echebarria04}, two dimensionless variables are 
now introduced. The first,
\be
u = \frac{\mu - \mu_{\rm eq}}{\kT}
\label{eq:udef}
\ee
is the dimensionless deviation of the chemical potential
from its equilibrium value. This variable can also be
expressed in terms of the composition and the
equilibrium composition at two-phase coexistence using 
the definition of Eq.~(\ref{eq:cprofile}) as
\be
u=\ln \frac{c(\phi,\mu)}{c_{\rm eq}(\phi)}.
\ee
The second dimensionless variable,
\be
U = \frac{e^u - 1}{1-k}
\label{eq:Udef}
\ee
is a dimensionless supersaturation. This can be seen by
inserting the identity $e^u=c(\phi,\mu)/c_{\rm eq}(\phi)$
obtained from the preceding equation, which yields
\be
U = \frac{c(\phi,\mu)-c_{\rm eq}(\phi)}{(1-k)c_{\rm eq}(\phi)}.
\label{eq:Uofceq}
\ee
When the driving force (the grand potential difference) is 
expressed in either of these variables, the same evolution equation 
for the phase field as in Ref.~\cite{Echebarria04} is obtained.

Matters are slightly more complicated for the second evolution
equation. Ref.~\cite{Echebarria04} uses an evolution equation for the 
composition (or, equivalently, for the dimensionless supersaturation $U$)
rather than for the chemical potential. For the dilute alloy model,
such an equation can be obtained starting from the general 
evolution equation of the chemical potential, 
Eq.~(\ref{eq:muevolution}), or (more simply) from the 
original formulation of the mass conservation law, 
Eq.~(\ref{eq:massconservation}). For this purpose, the
chemical potential has to be eliminated in favor of $c$
or $U$. This is possible because Eqs.~(\ref{eq:udef}) and
Eq.~(\ref{eq:Udef}) can be inverted and combined with
Eq.~(\ref{eq:csimplified}) to yield
\be
\mu = \mu_{\rm eq} + \kT\ln\frac{c}{c_l^{\rm eq}\left[1-(1-k)g_s(\phi)\right]}
\ee
and
\be
c = c_l^{\rm eq}\left[1-(1-k)g_s(\phi)\right]\left[1+(1-k)U\right].
\ee
A straightforward calculation then yields the variational
form (without the antitrapping current) of the evolution
equation for the composition of Ref.~\cite{Echebarria04}. 
These steps will be discussed in more details later on; here, it 
is important to stress that the possibility to switch from an evolution 
equation for $\mu$ to one for $c$ by an exact transformation 
is specific to the dilute alloy model: this procedure only works 
because the function $c(\phi,\mu)$ is easily inverted, which is 
not the case in more general models, as will be seen below.

\subsection{Ideal solution model}

In an ideal solution model, the free energy is a weighted
average of the pure substance free energies of A and B
plus an entropy of mixing term,
\be
f_{\nu}(T,c)= (1-c)f_{\nu}^A(T)+cf_{\nu}^B(T)+\frac{\kT}{\Va}
  \left[c\ln c + (1-c)\ln(1-c)\right].
\label{eq:fideal}
\ee
Contact with the notations of the previous examples can be
made by setting $\epsilon_{\nu}=f_{\nu}^B-f_{\nu}^A$; the
free energy becomes
\be
f_{\nu}(T,c)= f_{\nu}^A(T)+c\epsilon_{\nu}(T)+\frac{\kT}{\Va}
  \left[c\ln c + (1-c)\ln(1-c)\right].
\ee
The chemical potential is
\be
\mu=\frac{\partial f_{\nu}}{\partial \rho} = \Va\epsilon_{\nu}+\kT\ln\frac{c}{1-c}.
\ee
This can be inverted to yield the conentration in each phase
as a function of $\mu$,
\be
c_{\nu}(\mu) = \frac{\exp\left(\frac{\mu - \Va\epsilon_{\nu}}{\kT}\right)}
                {1+\exp\left(\frac{\mu - \Va\epsilon_{\nu}}{\kT}\right)}.
\ee
The grand potential densities are
\be
\omega_{s,l} = f_{\nu}^A + \frac{\kT}{\Va}\ln(1-c) 
             = f_{\nu}^A - \frac{\kT}{\Va}
             \ln\left[1+\exp\left(\frac{\mu - \Va\epsilon_{\nu}}{\kT}\right)\right].
\label{eq:omegadensitiesideal}
\ee
As before, these expressions can now be used to define the
interpolated composition and the susceptibility,
\be
c(\phi,\mu)=\frac{\exp\left(\frac{\mu - \Va\epsilon_s}{\kT}\right)}
                {1+\exp\left(\frac{\mu - \Va\epsilon_s}{\kT}\right)}g_s(\phi) +
              \frac{\exp\left(\frac{\mu - \Va\epsilon_l}{\kT}\right)}
                {1+\exp\left(\frac{\mu - \Va\epsilon_l}{\kT}\right)}(1-g_s(\phi)),
\label{eq:cofphiandmuideal}
\ee
\be
\chi(\phi,\mu) =  \frac{c_s(\mu)(1-c_s(\mu))g_s(\phi)+c_l(\mu)(1-c_l(\mu))(1-g_s(\phi))}{\Va\kT},
\ee
where the latter has been expressed in terms of the functions
$c_{s,l}(\mu)$ because this leads to a simpler expression.

The equations of motion for the ideal solution model are obtained
by inserting these expressions in the general evolution equations,
Eqs.~(\ref{eq:phievolution}) and (\ref{eq:muevolution}). The
equilibrium chemical potential and the phase diagram can be
calculated analytically, but the resulting expressions are
quite complicated and not of interest here. The
important point here is that for this model
it is not possible to transform the evolution equation
for $\mu$ into one for the composition $c$: whereas the 
functions $c_{\nu}(\mu)$ for the composition in
each phase as a function of $\mu$ can be easily inverted,
the same is not true for the interpolated composition given by
Eq.~(\ref{eq:cofphiandmuideal}). As a consequence, it is easy
to obtain $c$ from $\mu$ for given $\phi$ and $T$, but hard
to obtain $\mu$ from $c$.

\section{Relations with other phase-field models}
\label{sec_relations}

\subsection{Equivalence to the Kim-Kim-Suzuki model}

In the two-phase model \cite{Kim99}, the interface region is seen 
as a phenomenological superposition of the two bulk phases,
with a weighting function $h_s(\phi)$ that interpolates between
liquid ($h_s=0$) and solid ($h_s=1$). The main difference to
the coarse-graining approach is that the two-phase model
starts with two separate composition fields for the
solid and the liquid, $c_s$ and $c_l$. The free energy
density and the ``true'' composition (in the coarse-graining
sense) are then written as
\be
f(\phi,c) = h_s(\phi)f_s(c_s) + (1-h_s(\phi))f_l(c_l),
\ee
\be
c = h_s(\phi) c_s + (1-h_s(\phi)) c_l.
\label{eq:cdefmixture}
\ee
Since there are two equations but three variables ($\phi$, 
$c_s$, and $c_l$), an additional condition is needed to
close the system: the chemical potentials of the two 
coexisting phases are required to be the same,
\be
\mu \equiv \frac{\partial f_s(\phi,c_s)}{\partial c_s} =
           \frac{\partial f_l(\phi,c_l)}{\partial c_l}.
\label{eq:muequal}
\ee
Taking this implicit relationship between $c_s$ and $c_l$
into account, the number of independent fields is reduced
to two, and two evolution equations for, say, $\phi$ and
$c$ can be written down.

This approach is completely equivalent to the grand-potential
formalism outlined above, with the difference that the dynamical
variable is $c$ instead of $\mu$. To see this, set 
$h_s(\phi)=g_s(\phi)$. Equation (\ref{eq:cdefmixture})
becomes identical to Eq.~(\ref{eq:cofmuandphi}). Furthermore,
since the compositions $c_s$ and $c_l$ in Eq.~(\ref{eq:cofmuandphi})
are defined as functions of the variable $\mu$, Eq.~(\ref{eq:muequal})
is automatically satisfied. Finally, since all the equations
of Ref.~\cite{Kim99} are developed by analogy with the pure substance
model, it is not surprising that the evolution equations,
Eqs.~(31) and (32) of Ref.~\cite{Kim99} are identical to the
evolution equations (\ref{eq:phievolution}) and 
(\ref{eq:muevolution}) of the present paper
once all the notations have been properly translated: the
driving force for the phase transformation is the difference
in grand potential density, and the quantity $f_{cc}$ that
appears in the evolution equation for the concentration in
Ref.~\cite{Kim99} is identical to $1/(\Va^2\chi)$ here.

In summary, the two-phase model can be obtained in a fully
variational manner from a grand-potential functional. Note 
that, in this point of view, the introduction of two separate
composition fields is not necessary any more: the fundamental
dynamical field is the chemical potential, and the two fields
$c_s$ and $c_l$ are simply obtained as the derivatives of the
bulk grand potential densities with respect to $\mu$, whereas 
the ``real'' composition at any space point is given by the
functional derivative of the grand-potential functional.

\subsection{Relation to the phenomenological two-phase model}

At this point, it is useful to discuss the respective merits 
of the grand-potential and the two-phase approaches.
At first glance, the former seems to be more advantageous: 
the number of fundamental fields is equal to two (as compared 
to three for the two-phase model), and whereas $\mu$ has to be 
obtained from $c_s$ and $c_l$ in the two-phase model by solving 
Eq.~(\ref{eq:muequal}), $c_s$ and $c_l$ are obtained from $\mu$
by simple derivatives in the grand-potential formalism. However, 
an analysis in terms of computation time reveals that matters
are not so simple. For the sake of concreteness, consider the 
simplest case of constant but distinct solute diffusivities in 
the solid and the liquid. In this case, the evolution equation
for the composition reduces, in the bulk phases, to the
simple linear diffusion equation. In contrast, the evolution
equation for $\mu$ has several nonlinearities (even in the
bulk) due to the presence of the factors $\chi(\phi,\mu)$,
which are in general nonlinear functions of $\mu$. This means 
that the numerical effort to integrate the equation for $\mu$ 
in the bulk is higher that the one for integrating an equation 
for $c$. In the interfaces, the grand potential formulation 
does not require much additional effort, whereas the nontrivial 
Eq.~(\ref{eq:muequal}) has to be solved in the two-phase model.
However, since the interface regions usually represent only a 
small fraction of all the grid points in a numerical simulation,
the equation of motion for $\mu$ may not always be advantageous 
from a computational point of view.

The computational disadvantage of the grand-potential formalism
in the bulk can be alleviated by an additional step, which also
brings to light the direct relation of this approach to the
phenomenological two-phase model developed by
the Access group \cite{Tiaden98}. The
idea is to make a change of variables and to replace the
chemical potential field by another continous field that plays
the same role. In order to obtain the standard diffusion
equation for this field in the liquid phase, the appropriate
field is the density (or the composition) in the liquid,
$\rho_l$ or $c_l$. Indeed, under the hypothesis that
$\mu = \partial f_l/\partial\rho$ is an invertible function
of $\rho$, the function $\rho_l = \partial\omega_l/\partial\mu$
is just its inverse function, according to the properties of
the Legendre transform. Furthermore, 
$\rho_s = \partial\omega_s/\partial\mu$ is also a function of
$\mu$ and can therefore expressed as a function of $\rho_l$,
\begin{equation}
\rho_s(\rho_l) = \rho_s(\mu(\rho_l)).
\end{equation}
Eliminating $\mu$ in favor of $\rho_l$ in Eq. (\ref{eq:muevolution})
yields
\begin{equation}
\partial_t \rho_l = \frac{\chi_l(\rho_l)}{\chi(\phi,\rho_l)}
\left\{\vec\nabla\cdot\left[D(\phi,\rho_l)\frac{\chi(\phi,\rho_l)}{\chi_l(\rho_l)}\vec\nabla\rho_l\right] +
      \frac{g'(\phi)}{2}\left[\rho_s(\rho_l)-\rho_l\right]\partial_t\phi\right\},
\end{equation}
where $\chi_l=\partial^2 \omega_l/\partial\mu^2=\chi(-1,\mu)$ is
the susceptibility of the liquid phase. Unsing the fact that
$\chi(\phi,\mu)=\chi_s(\mu)g_s(\phi)+\chi_l(\mu)(1-g_s(\phi))$
by definition, this equation can be rewritten as
\begin{eqnarray}
\partial_t \rho_l & = & \frac{1}{\chi_l(\rho_l)[1-(1-g_s(\phi)\chi_s/\chi_l)]} \times
\nonumber \\
 & & 
\biggl\{\vec\nabla\cdot\left[D(\phi,\rho_l)\chi_l(\rho_l)[1-(1-g_s(\phi)\chi_s/\chi_l)]\vec\nabla\rho_l\right] + \nonumber \\
 & & \quad
      \frac{g'(\phi)}{2}\left[\rho_s(\rho_l)-\rho_l\right]\partial_t\phi\biggr\}.
\end{eqnarray}
For a dilute alloy, $\rho_s/\rho_l=k$ and $\chi_s/\chi_l=k$ according
to Eqs. (\ref{eq:kdef}) and (\ref{eq:chidilute}), and the above
model becomes identical to the one of Ref.~\cite{Tiaden98}. The
grand-potential formalism thus allows for a generalization of
this model to arbitrary phase diagrams.

\subsection{Local supersaturation approximation}

Further progress can be made by a simple approximation.
The crucial point is the relation between 
composition and chemical potential: its nonlinearity 
penalizes the grand-potential formalism in the bulk and makes 
the resolution of Eq.~(\ref{eq:muequal}) in the two-phase model 
non-trivial. The idea is thus to replace the exact relation 
between chemical potential and 
composition {\em in the interfaces} by an approximate one 
that will make it possible to write a simple equation for
the composition. This approximation will be called {\em local 
supersaturation approximation}: it exploits the fact that,
for slow solidification, the chemical potential in the interfaces 
is close to the value for two-phase equilibrium. This suggests to 
use a Taylor expansion of the composition around the equilibrium
composition profile in the interface,
\be
c - c_{\rm eq}(\phi) = \left.\frac{\partial c}{\partial \mu}\right|_{\mu_{\rm eq}}(\mu-\mu_{\rm eq})
 = \Va\chi(\phi,\mu_{\rm eq})(\mu-\mu_{\rm eq})
\label{eq:LSA} 
\ee
which can be easily inverted to yield
\be
\mu = \mu_{\rm eq} + \frac{c-c_{\rm eq}(\phi)}{\Va\chi(\phi,\mu_{\rm eq})}.
\ee
Furthermore, an expansion of the driving force around the
equilibrium chemical potential yields
\be
\omega_s(\mu) - \omega_l(\mu) = \frac{c_l^{\rm eq}-c_s^{\rm eq}}{\Va}(\mu-\mu_{\rm eq}).
\ee
Inserting these expression in the evolution equation for
the phase field and the mass conservation law yields a
simple set of equations for $\phi$ and $c$,
\be
\frac{1}{M_\phi}\partial_t\phi = \sigma\vec\nabla^2\phi-Hf'_{\rm dw}
   -\frac{g'(\phi)}{2}\Delta c \frac{c-c_{\rm eq}(\phi)}{\Va^2\chi(\phi,\mu_{\rm eq})}
\label{eq:phiLSA}
\ee
\be
\partial_t c = \vec\nabla\left[D(\phi)\chi(\phi,\mu_{\rm eq})
  \vec\nabla\left(\frac{c-c_{\rm eq}(\phi)}{\chi(\phi,\mu_{\rm eq})}\right)\right].
\label{eq:cLSA}
\ee
Note that the second equation can also be rewritten after
applying the chain rule as
\be
\partial_t c = \vec\nabla D(\phi) \vec\nabla c
      + \vec\nabla \left[ D(\phi)\frac{g'(\phi)}{2}
       \left( \Delta c + \frac{\chi_l^{\rm eq}-\chi_s^{\rm eq}}
                              {\chi(\phi,\mu_{\rm eq})}\right)
                                 \vec\nabla\phi\right],
\ee
where $\chi_{s,l}^{\rm eq}=\chi(\pm 1,\mu_{\rm eq})$. This latter
form displays explicitly the two driving forces for solute 
diffusion that establish the equilibrium solute profile
and that lead to solute redistribution out of equilibrium: 
composition gradients and differences between the 
thermodynamic potentials in the two phases.

Before proceeding further, it is useful to relate the quantities 
that appear in the above equations to the phase diagram of the 
binary alloy, characterized by the curves $c_s^{\rm eq}(T)$ and
$c_l^{\rm eq}(T)$, or equivalently by the coexistence line $\mu_{\rm eq}(T)$.
The quantities $\chi_{\nu}^{\rm eq}$ are related to the liquidus
and solidus slopes, $m_{\nu}=dT/dc_{\nu}^{\rm eq}$,
\be
\frac {1}{m_{\nu}} = \frac{d c_{\nu}^{\rm eq}}{dT} = 
      \left.\frac{\partial c}{\partial \mu}\right|_{\mu_{\rm eq}}\frac{d\mu_{\rm eq}}{dT}
      = \Va\chi_{\nu}^{\rm eq}\frac{d\mu_{\rm eq}}{dT}.
\ee
Moreover, the quantity $d\mu_{\rm eq}/dT$ can be evaluated using
a Clausius-Clapeyron relation for the $\mu$-$T$ coexistence line,
\be
\frac{d\mu_{\rm eq}}{dT} = - \frac{L}{T\Delta\rho} = - \frac{L\Va}{T\Delta c}.
\ee
Combining these results yields
\be
\chi_{s,l}^{\rm eq} = -\frac{T\Delta c}{\Va^2Lm_{s,l}}.
\ee
Thus, the susceptibilities are
inversely proportional to the liquidus and solidus slopes,
and therefore $\chi_s^{\rm eq}/\chi_l^{\rm eq}=m_l/m_s$. Defining an effective 
partition coefficient by the ratio of the liquidus solpes $k_m=m_l/m_s$, 
the evolution equations of the phase-field model can be further
simplified. Indeed, the susceptibility along the equilibrium
profile is
\be
\chi(\phi,\mu_{\rm eq}) = \chi_s^{\rm eq}g_s(\phi) + \chi_l^{\rm eq}(1-g_s(\phi))
   = \chi_l^{\rm eq}[1-(1-k_m)g_s(\phi)],
\ee
and the evolution equations for $\phi$ and $c$ become
\be
\frac{1}{M_\phi}\partial_t\phi = \sigma\vec\nabla^2\phi-Hf'_{\rm dw}
   -\frac{g'(\phi)}{2}\frac{\Delta c}{\Va^2\chi_l^{\rm eq}} 
      \frac{c-c_{\rm eq}(\phi)}{1-(1-k_m)g_s(\phi)}
\label{eq:phiLSAkm}
\ee
\be
\partial_t c = \vec\nabla\left[D(\phi)[1-(1-k_m)g_s(\phi)]
  \vec\nabla\left(\frac{c-c_{\rm eq}(\phi)}{1-(1-k_m)g_s(\phi)}\right)\right].
\label{eq:cLSAkm}
\ee
These equations are very similar to the ones of the dilute alloy 
model, except that the partition coefficient $k$ has been replaced
by the effective partition coefficient $k_m$ which depends
on the temperature. In that sense, this approach bears some
similarity with the method used in Ref.~\cite{Tong08} to construct
a quantitative phase-field model for arbitrary phase diagrams:
the free energy curves are first approximated by a dilute
alloy phase diagram with ``effective'' (temperature-dependent)
partition coefficient, melting temperature, and liquidus slope;
the equations of motion for the dilute alloy model are then
applied with these effective parameters. Here, the 
approximation is directly in the
evolution equations, and can be applied in a straightforward
manner for arbitrary free energy functions. Also note that the
two appromixations are not completely equivalent. For instance,
the effective partition coefficient in Ref.~\cite{Tong08} is defined
by the ratio of the compositions, $c_s^{\rm eq}/c_l^{\rm eq}$, which
is equal to the ratio $m_l/m_s$ only for a dilute alloy.

It should also be noted that this approximation is {\em not}
equivalent to the approximation of constant concentration 
gap $\Delta c$ (equivalent to parallel liquidus and solidus 
slopes) used in Refs.~\cite{Losert98,Folch03,Folch05}. This can be 
easily seen from Eq.~(\ref{eq:LSA}) by considering a constant chemical
potential deviation $\delta\mu = \mu-\mu_{\rm eq}$ (generated, for example,
by a local curvature of the interface): the shifts in concentration
on the two sides of the interface are proportional to $\chi_{s,l}^{\rm eq}$,
respectively, and hence inversely proportional to the liquidus
slopes, as they should be.

The above approximation has been called ``local'' for two
reasons. First, this emphasizes the fact that the approximation
of the relationship between composition and chemical potential is 
needed only in the interfaces, while it leaves the bulk evolution 
equations unchanged. Second, it is anticipated that this method 
should be applicable to situations in which the temperature 
field varies slowly with time and over large
length scales, such as in directional solidification or in 
thermosolutal models with realistic values of the Lewis
number. For this purpose, it should be sufficient to apply the
above equations with the {\em local} value of the temperature
for each point of the interface, which implies that $\Delta c$,
$\chi_{s,l}^{\rm eq}$ and $k_m$ are not constants but vary between
different interface points.

\subsection{Non-variational model and antitrapping current}

The next step is to incorporate into the model two features
that have been widely used to increase the precision and
performance of phase-field models. The first is motivated
by the fact that a
non-variational model can be more efficient for computational 
purposes than a strictly variational formulation. This
was first highlighted by Karma and Rappel \cite{Karma98}, and their
method has since been used in many other models. In the
terms of the present formulation, it amounts to keeping
the interpolation function $g(\phi)$ in the evolution
equation of the phase field, but using a different
interpolation function for the number density and
the susceptibility. Let $h(\phi)$ be a function that has 
the property $h(\pm 1)=\pm1$, and let
\be
h_s(\phi) = \frac{1 + h(\phi)}{2}.
\ee
Then, the concentration is defined as
\be
\rho(\phi,\mu)=\frac{c(\phi,\mu)}{\Va} = \rho_s(\mu)h_s(\phi) + \rho_l(\mu)(1-h_s(\phi))
\label{eq:credef}
\ee
instead of Eq.~(\ref{eq:rhoofmuandphi}); as a consequence,
the equilibrium composition profile given by Eq.~(\ref{eq:cprofile}) 
is also modified and becomes
\be
c_{\rm eq}(\phi) = c_s^{\rm eq}h_s(\phi) + c_l^{\rm eq}(1-h_s(\phi)).
\ee
The susceptibility is still defined as the derivative of the
number density with respect to the chemical potential and
becomes
\be
\chi(\phi,T,\mu)=\frac{\partial\rho(\phi,T,\mu)}{\partial\mu}=
  h_s(\phi) \chi_s(T,\mu) + (1-h_s(\phi)) \chi_l(T,\mu).
\label{eq:chiredef}
\ee

The second feature is the so-called antitrapping current,
which was developed in order to counterbalance spurious solute 
trapping \cite{Karma01,Echebarria04}. Indeed, if the solute diffusivity 
in the solid is substantially lower than in the liquid, as is
usually the case in alloy solidification, the upscaling
of the interface thickness magnifies the solute trapping
effect, whose magnitude is proportional to the interface
thickness \cite{Aziz82}. To restore local equilibrium at the interface,
as appropriate for low-speed solidification, an additional
solute current is introduced which ``pushes'' the solute
out of the freezing material, and which is given by
\be
\vec j_{\rm at} = aW[\rho_l(\mu)-\rho_s(\mu)]\hat n \partial_t\phi,
\ee
where $\hat n = -\vec\nabla\phi/|\vec\nabla\phi|$ is the
unit normal vector pointing from the solid to the liquid,
$W$ is the interface thickness, and $a>0$ is a constant to be 
determined by a matched asymptotic analysis \cite{Echebarria04}. 
The current thus defined is proportional 
to the interface velocity ({\em via} the factor $\partial_t\phi$)
and to the composition difference between solid and liquid,
and is directed from the solid to the liquid for a solidifying
interface, for which $\partial_t\phi >0$. The mass conservation
law, Eq.~(\ref{eq:massconservation}), reads now
\be
\partial_t \rho = -\vec\nabla \left( \vec j_\rho + \vec j_{\rm at}\right) =
                  \vec\nabla \left( M(\phi,\mu)\vec\nabla\mu -
                   aW[\rho_l(\mu)-\rho_s(\mu)]\hat n \partial_t\phi\right).
\ee

Taking these modifications into account, the evolution equation
for the chemical potential, Eq.~(\ref{eq:muevolution}), is replaced by
\begin{eqnarray}
\partial_t\mu & = & \frac{1}{\chi(\phi,\mu)}
   \biggl\{\vec\nabla\cdot\left[D(\phi,\mu)\chi(\phi,\mu)\vec\nabla\mu
   - aW[\rho_l(\mu)-\rho_s(\mu)]\hat n \partial_t\phi\right] \nonumber \\
   & & \qquad\qquad\quad \mbox{} -
      \frac{h'(\phi)}{2}\left[\rho_s(\mu)-\rho_l(\mu)\right]\partial_t\phi\biggr\};
\end{eqnarray}
the evolution equation for $\phi$ remains unchanged.

In the local supersaturation approximation, the composition
difference in the expression for the antitrapping current
is approximated by its equilibrium value, 
$\rho_l-\rho_s=(c_l^{\rm eq}-c_s^{\rm eq})/\Va = \Delta c/\Va$.
The evolution equation for the concentration,
Eq.~(\ref{eq:cLSAkm}), is then replaced by
\be
\partial_t c = \vec\nabla\left[D(\phi)[1-(1-k_m)h_s(\phi)]
  \vec\nabla\left(\frac{c-c_{\rm eq}(\phi)}{1-(1-k_m)h_s(\phi)}\right)
   - aW\hat n\Delta c \partial_t\phi\right].
\label{eq:cLSAkmredef}
\ee

\subsection{Relation to the quantitative dilute alloy model}
As usual, the parameters of the phase-field model have to
be related to the quantities that appear in the sharp-interface
theories by matched asymptotic analysis. The complete asymptotic
analysis for the general Eqs. (\ref{eq:phievolution}) and
(\ref{eq:muevolution}) will be presented elsewhere. Here, only the
behavior of the model in the local supersaturation approximation
will be discussed, because for its analysis the analogy to the
dilute alloy model \cite{Echebarria04} can be exploited.

In order to apply directly the results of Ref.~\cite{Echebarria04}, 
it is useful to cast the evolution equations in dimensionless
form. From Eqs.~(\ref{eq:phiLSA}) and (\ref{eq:cLSA}), it is
clear that the dimensionless variable that generalizes the
quantity $U$ of the dilute alloy model defined by 
Eq.~(\ref{eq:Udef}) is
\be
U = \frac{\chi_l^{\rm eq}}{\chi(\phi,\mu_{\rm eq})}\frac{c-c_{\rm eq}(\phi)}{\Delta c} = 
    \frac{c-c_{\rm eq}(\phi)}{\Delta c[1-(1-k_m)h_s(\phi)]}.
\ee
Indeed, this expression can be rewritten using the fact
that $c_{\rm eq}(\phi)=c_l^{\rm eq}[1-(1-k)h_s(\phi)]$, where
$k=c_s^{\rm eq}/c_l^{\rm eq}$ is the standard partition coefficient,
\be
U = \frac{1}{1-k} \frac{1-(1-k)h_s(\phi)}{1-(1-k_m)h_s(\phi)}
    \frac{c-c_{\rm eq}(\phi)}{c_{\rm eq}(\phi)},
\ee
which reduces to Eq.~(\ref{eq:Uofceq}) for a dilute alloy, since
$k=k_m$ in this case.

In terms of this variable, Eq.~(\ref{eq:cLSA}) becomes
\begin{eqnarray}
[1-(1-k_m)h_s(\phi)]\partial_t U & = & \vec\nabla\bigl\{
  D(\phi)[1-(1-k_m)h_s(\phi)] \vec\nabla U  \nonumber \\
  & &  \quad \mbox{}+ aW\hat n[1+(1-k_m)U]\partial_t\phi\bigr\}
 \nonumber \\
 & & \mbox{}+ [1+(1-k_m)U]\partial_t h_s(\phi).
\end{eqnarray}
A form formally identical to Eq.~(69) in Ref.~\cite{Echebarria04} is 
obtained by choosing a particular interpolation for the diffusion 
coefficient $D(\phi)$, namely by setting
\be
D(\phi)[1-(1-k_m)h_s(\phi)] \equiv D_l q(\phi),
\ee
where $D_l$ is the solute diffusivity in the liquid, supposed to
be constant. Note that, since the left hand side is actually the 
product of the diffusivity and the susceptibility, strictly speaking 
the function $q(\phi)$ is an interpolation of the {\em mobility}
rather than of the diffusion coefficient. The final result
for the evolution equation is
\begin{eqnarray}
[1-(1-k_m)h_s(\phi)]\partial_t U & = & \vec\nabla\left(
  D_l q(\phi) \vec\nabla U + aW\hat n[1+(1-k_m)U]\partial_t\phi\right)
 \nonumber \\
 & & \mbox{}+ [1+(1-k_m)U]\partial_t h_s(\phi).
\label{eq:Ulikedilute}
\end{eqnarray}
This equation is indeed identical to the one used in the asymptotics
of the dilute alloy model, except that the dilute alloy partition
coefficient $k$ is replaced by the ratio of the solidus and liquidus
slopes $k_m$. This is very natural, since this quantity controls how
the composition difference between solid and liquid depends on the
chemical potential at the interface.

The replacement of $c$ by $U$ in the evolution equation for the
phase field leads to
\be
\frac{1}{M_\phi}\partial_t\phi = \sigma\vec\nabla^2\phi-Hf'_{\rm dw}
   -\frac{g'(\phi)}{2}\frac{(\Delta c)^2}{\Va\chi_l^{\rm eq}} U.
\ee
This equation is now divided by the constant $H$, which amounts
to non-dimen\-sionalizing the free energy and grand potential
densities (since $H$ has dimension of energy per unit volume).
Furthermore, the function $g$ is now chosen to be the standard
fifth-order polynomial $g(\phi)=15(\phi-2\phi^3/3+\phi^5/5)/8$,
the function $h(\phi)=\phi$, and the double-well function to 
be $f_{\rm dw}=1/4-\phi^2/2+\phi^4/4$. The resulting equation reads
\be
\tau\partial_t\phi = W^2\vec\nabla^2\phi + \phi-\phi^3
   -(1-\phi^2)^2\lambda U,
\label{eq:phinodim}
\ee
where $\tau=1/(M_\phi H)$ is the relaxation time for the phase
field, $W=\sqrt{\sigma/H}$ is the interface thickness defined 
by Eq.~(\ref{eq:Wdef}), and the constant $\lambda$ is given by
\be
\lambda = \frac{15}{16} \frac{(\Delta c)^2}{H\Va^2\chi_l^{\rm eq}}.
\ee
Equation (\ref{eq:phinodim}) is identical to the standard evolution
equation for the phase field \cite{Karma98,Echebarria04}.

As announced previously, the results of the asymptotic analysis of 
Ref.~\cite{Echebarria04} can now be exploited since 
Eqs.~(\ref{eq:Ulikedilute}) and
(\ref{eq:phinodim}) are identically to the model analyzed in this
reference. Therefore, the variable $U$ obeys, in the liquid,
the free boundary problem
\be
\partial_t U = D_l\vec\nabla^2 U,
\ee
\be
U_{\rm int} = - d_0{\cal K} - \beta V_n,
\label{eq:UGiTho}
\ee
\be
[1+(1-k_m)U_{\rm int}] V_n = - D_l\partial_n U|_{\rm int},
\label{eq:UStefan}
\ee 
where ${\cal K}$ and $V_n$ are the local curvature and interface
velocity, respectively, $d_0$ is the capillary length, and
$\beta$ the kinetic coefficient. Equation (\ref{eq:UGiTho})
is the generalized Gibbs-Thomson equation, and Eq.~(\ref{eq:UStefan})
is the Stefan boundary condition that describes mass conservation
at the phase boudary.

In terms of the phase-field parameters, the capillary length and
the kinetic coefficient are given by
\be
d_0 = a_1 \frac{W}{\lambda}
\label{eq:d0defPF}
\ee
\be
\beta = a_1\frac{\tau}{\lambda W}\left(1-a_2\frac{\lambda W^2}{\tau D_l}\right)
\ee
with $a_1=5\sqrt{2}/8$ and $a_2=0.6267$; these values are identical 
to those obtained by Karma and Rappel \cite{Karma98}.

In terms of physical quantites, this expression for the capillary 
length is in fact identical to the standard thermodynamic 
definition \cite{LangerBegRohu}.
Indeed, the number $a_1=5\sqrt{2}/8$ quoted above is equal to
$(15/16)I$, where $I$, the numerical constant defined in 
Eq. (\ref{eq:gammarelation}), is equal to $2\sqrt{2}/3$ for
the standard fourth-order double well potential used here.
With the help of these relations, Eq. (\ref{eq:d0defPF}) can
be rewritten as
\be
d_0 = \frac{IWH \Va^2\chi_l^{\rm eq}}{(\Delta c)^2}.
\ee
Then, the use of Eqs. (\ref{eq:gammarelation}) and (\ref{eq:Darken})
yields
\be
d_0 = \frac{\gamma\Va^2\chi_l^{\rm eq}}{(\Delta c)^2}= 
      \frac{\gamma}{(\Delta c)^2 
         \left.\frac{\partial^2f_l(c)}{\partial c^2}\right|_{c_l}}.
\label{eq:d0def}
\ee

\section{Numerical tests}
\label{sec_numerics}
The relations found in the preceding section are now used to
perform some illustrative simulations on a concrete model system.
The ideal solution model is chosen, with the same parameters as
those used in Ref.~\cite{Warren95} to model the Nickel-Copper
alloy. This alloy exhibits a typical lens-shape phase diagram
with a rather narrow coexistence zone. Concretely, the free
energy densities given by Eq. (\ref{eq:fideal}) are used; the
free energy differences (between solid and liquid) of the pure 
substances are given by
\be
f_s^{Ni}(T)-f_l^{Ni}(T)=\frac{L^{Ni}}{T_m^{Ni}}(T-T_m^{Ni}),
\ee
\be
f_s^{Cu}(T)-f_l^{Cu}(T)=\frac{L^{Cu}}{T_m^{Cu}}(T-T_m^{Ni}),
\ee
with the melting temperatures $T_m^{Ni}=1728$ K and $T_m^{Cu}=1358$ K
and the latent heats $L^{Ni}=2350$ J/cm$^3$ and $L^{Cu}=1728$ J/cm$^3$.
The molar volume is taken to be $7.42$ cm$^3$, and the surface tension
$\gamma=3.3\times 10^{-5}$ J/cm$^2$.

For a temperature of $1710$ K, the equilibrium concentrations are 
$c_s^{\rm eq}=0.045988$ and $c_l^{\rm eq}=0.058098$, the partition
coefficient is $k=0.7916$, the ratio of the liquidus slopes is $k_m=0.8017$,
and the capillary length calculated by Eq.~(\ref{eq:d0def}) is 
$d_0=6,426\times 10^{-6}$ cm. Isothermal dendritic solidification is
simulated in two dimensions. The anisotropy needed to obtain stable
dendritic growth is introduced in the standard way 
\cite{Karma98,Karma01,Echebarria04} by making the surface tension
dependent on the angle $\theta$ between the interface normal and
a crystallographic axis, here chosen to coincide with the $x$ 
direction. A standard fourfold anisotropy, 
$\gamma(\theta)=\bar\gamma(1+\epsilon_4\cos(4\theta))$ is used,
with $\epsilon_4=0.025$. The initial composition of the liquid
is chosen as $(c_s^{\rm eq}+c_l^{\rm eq})/2$, which corresponds to a
supersaturation of $0.5$.

\begin{figure}
\centering
\includegraphics[width=.8\textwidth]{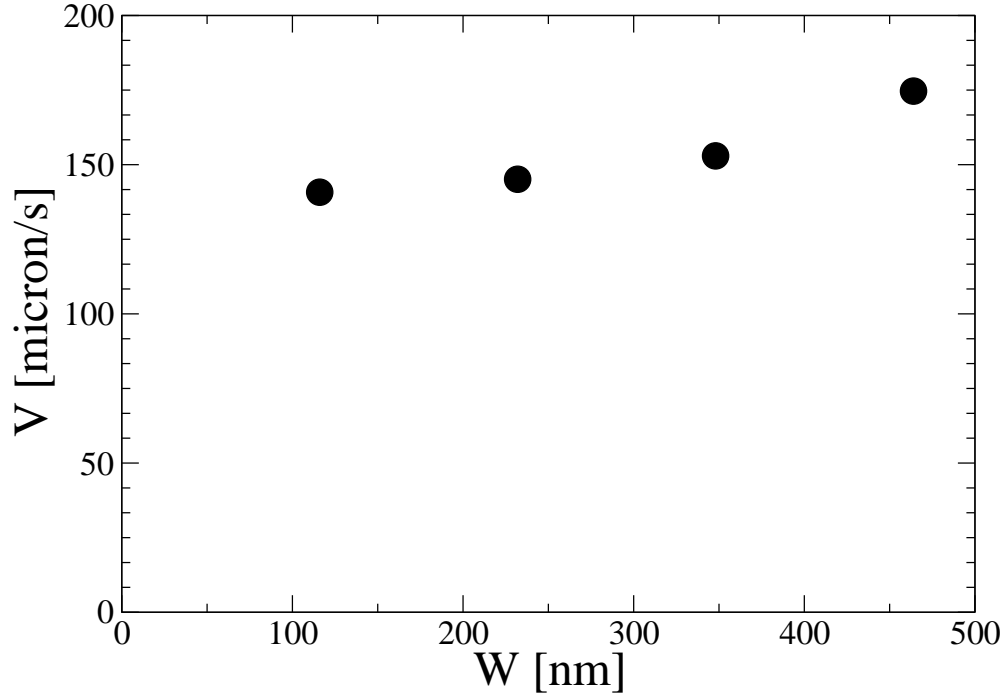}
\caption{Growth velocity of a two-dimensional dendrite in the
Ni-Cu alloy at $1710$ K, for different choices of the interface
width $W$.}
\label{figvel}
\end{figure}

Four simulations are carried out with the equations of motion
in the local supersaturation approximation, with $\lambda=1.596,
3.192, 4.788$, and $6.384$, which corresponds to interface widths 
of $W=116, 232, 348,$ and $464$ nm. For each simulation, the
relaxation time of the phase-field equation is chosen such as
to eliminate interface kinetics ($\beta=0$ for all orientations).
The steady-state growth velocity of the dendrites is measured
as described in Ref.~\cite{Karma98}, and the result is displayed 
in Fig. \ref{figvel}. It exhibits the behavior that is typical
for quantitative phase-field models: the simulation results
converge to a constant value with decreasing interface thickness
$W$, and the convergence is roughly quadratic in $W$, which can
be expected since all terms linear in $W$ have been eliminated
by the model formulation. It is not surprising to find such behavior, 
since the model used here is essentially identical to the
dilute alloy model for which rapid convergence with $W$ has been
demonstrated \cite{Karma01,Echebarria04}. Moreover, the parameters
chosen here are in a region of the phase diagram in which the
dilute approximation should still work quite well. However, this
is not a limitation of the approach: simulations at $T=1410$ K,
where the ratio of the liquidus slopes $k_m=1.2609$ is very
different from the partition coefficient $k=0.9590$, yield
a similar convergence plot. This shows that the model can be 
applied to alloys with arbitrary phase diagram.

\section{Summary and Perspectives}
\label{sec_final}
The most important conclusions of the present work can be
summarized as follows.
\begin{enumerate}
\item A phase-field model for alloy solidification has been
obtained in a completely variational framework, starting
from a phenomenological grand-potential functional that is
a simple sum of bulk and interface contributions. In this
model, the two dynamic variables are the phase field and
the chemical potential field. A complete analogy can be established
with the standard phase-field model for the solidification of a 
pure substance, in which the variables are the phase field and
the temperature field. The main difference between the two sets of 
equations is that the relation between the composition and the 
chemical potential (the extensive and the intensive variable) is
nonlinear, whereas the relation between temperature and
internal energy is usually assumed to be linear.
\item The resulting model is shown to be completely equivalent 
to the model of Kim,Kim and Suzuki \cite{Kim99}, with the
chemical potential replacing the composition as the dynamic
field. As a result, in the present model no ``partitioning'' of
the solute into coexisting phases is necessary; the (auxiliary)
compositions in each of the phases can be simply obtained from
the chemical potential. With an additional change of variables,
the phenomenological Access model \cite{Tiaden98} can also be
recovered and extended to general alloy phase diagrams. These
developments show that these two-phase models, despite a seemingly 
quite different starting point, can in fact also be obtained by a
coarse-graining procedure if the appropriate thermodynamic
potential is used.
\item The equations of the new model can also be written in terms
of a phase field and the composition field. They thus have 
essentially the same computational complexity as the original
phase-field models for alloy solidification \cite{Wheeler92,Caginalp93}
derived in the coarse-graining framework. However, in 
contrast to the latter models, they have a decisive property 
which is required for quantitative simulations: bulk and interface
thermodynamic properties can be adjusted independently. This
difference in behavior is due to the different interpolations of
the relevant thermodynamic potentials (free energies in 
Refs. \cite{Wheeler92,Caginalp93}, grand potentials here).
\item With an additional approximation -- a linearization of the
relation between chemical potential and composition inside the
interfaces -- the model becomes equivalent to the quantitative
dilute alloy model studied in Refs.~\cite{Karma01,Echebarria04}. 
This feature makes it possible to include the additional antitrapping
current as in that model, and to apply the detailed asymptotic 
analysis of Ref.~\cite{Echebarria04}. As demonstrated for one 
particular case here, efficient and accurate simulations are 
thus possible for arbitrary alloy phase diagrams.
\end{enumerate}

Numerous interesting perpectives for future work arise from
the present results. First, the model has been worked out here for
isothermal solidification only, but it can be easily extended
to non-isothermal situations: coupled equations for the phase field,
the temperature field and the chemical potential field can be
developed by following the same steps as done here for each
transport field separately, taking also into account
off-diagonal elements in the Onsager matrix of transport
coefficients as well as cross-derivatives of the thermodynamic
potentials. This is anticipated to yield a generalization of the thermosolutal
model of Ref.~\cite{Ramirez04}.

Moreover, the generalization of the formalism to multi-component
systems should be straightforward. This is particularly interesting
because general models for multi-component multi-phase solidification
have been developed in the two-phase framework \cite{Eiken06,Kim07,Steinbach09}.
In these models, the determination of the compositions in the individual
phases for given global composition requires to solve a generalization of
Eq.~(\ref{eq:muequal}), which represents a set of coupled nonlinear 
equations (one for each component). A
formulation in terms of the chemical potential would completely
avoid this problem and thus potentially offer important gains in
computational performance.

It should also be mentioned that the two-phase approach has been
used in other contexts. One example is the
treatment of fluid flow in solidification \cite{Beckermann99} and 
two-phase flows \cite{Sun04,Sun08}, where two separate velocity fields,
one for each phase, are introduced, in contrast to
``coarse-graining'' models in which a single velocity field
is used \cite{Anderson00}. It would be interesting to reassess
the relations between these different models in the light of the
present findings.

\begin{acknowledgments}
I thank Abhik Choudhury, who has independently pursued research
along similar lines, for stimulating discussions.
\end{acknowledgments}


\end{document}